\documentclass[preprint,preprintnumbers,amsmath,amssymb]{revtex4}
\usepackage{graphicx} 
\usepackage{dcolumn} 
\usepackage{bm} 
\usepackage{natbib}

\begin{document}
\title{Muon-spin rotation measurements of the vortex state in Sr$_2$RuO$_4$:
type-1.5 superconductivity, vortex clustering and  a crossover from a triangular to a square vortex lattice}
\author{S.J. ~Ray$^{1}$,  A.S.~Gibbs$^{1}$,  S.J.~Bending$^{2}$,  P.J.~Curran$^{2}$, E.~Babaev$^{3,4}$, C.~Baines$^{5}$, A.P.~Mackenzie$^{1,6}$, and S.L.~Lee$^{1}$}
\affiliation{$^{1}$School of Physics and Astronomy, SUPA, University of St.Andrews, KY16 9SS, UK.\\
$^{2}$Department of Physics, University of Bath, Claverton Down, Bath, BA2 7AY, UK.\\
$^{3}$Department of Theoretical Physics, The Royal Institute of Technology, Stockholm, SE-10691 Sweden.\\
$^{4}$Department  of Physics, University of Massachusetts Amherst, MA 01003 USA. \\
$^{5}$Labor f\"ur Myonspinspektroskopie, Paul Scherrer Institute, CH-5232 Villigen PSI, Switzerland.\\
$^{6}$Max Planck Institute for Chemical Physics of Solids, Noethnitzer Str. 40, D-01187 Dresden, Germany.
}
\date{\today}
\begin{abstract}
Muon-spin rotation has been used to probe vortex state in  Sr$_2$RuO$_4$.  At moderate fields and temperatures a lattice of triangular symmetry is observed, crossing over to a lattice of square symmetry with increasing field and temperature. At lower fields it is found that there are large regions of the sample that are completely free from vortices which grow in volume  as the temperature falls.
Importantly this is accompanied by {\it increasing} vortex density and increasing disorder within the vortex-cluster containing regions.  Both effects are expected to result from the strongly temperature-dependent long-range vortex attractive forces arising from the multi-band chiral-order superconductivity. 
  \end{abstract}
\maketitle
\setcounter{figure}{0}

\section{Introduction}
The unconventional superconductor Sr$_2$RuO$_4$  is a highly two-dimensional layered perovskite  with  an in-plane superconducting coherence length $\xi_{ab}\sim 66$~nm and in the most pure samples a superconducting  transition temperature $T_c=1.5$~K  \cite{Andy-review}.
There is significant evidence to suggest that the system is a multiband superconductor \cite{Deguchi-2004b,Deguchi-2004a}.  Knight shift measurements provide convincing evidence for a triplet pairing state \cite{Ishida-98} and a number of experiments  indicate an odd-parity orbital symmetry state \cite{Mackenzie-98,Nelson-2004,Kapitulnik-2006}. Reports of spontaneous time reversal symmetry breaking (TRSB) {\em via}  zero-field  muon-spin rotation ($\mu$SR)  and  polar Kerr effect measurements  \cite{Luke-98,Kapitulnik-2006} support suggestions that the orbital order parameter is chiral $p$-wave of the form $p_x \pm ip_y$.  Scanning probe experiments have however failed to detect signatures of spontaneous magnetisation \cite{Cliff-2010,Bending-2011}, although  it has been argued that due to the multiband nature of the superconductivity the edge currents can be small \cite{Raghu-2010}.

The vortex lattice in Sr$_2$RuO$_4$ has also generated a number of interesting studies.  Neutron and muon measurements \cite{Riseman-98, Aegeter-98} first demonstrated that at fields down to 5 mT the vortex lattice was of square symmetry.  More recently a number of scanning probe measurements have been performed \cite{Dolocan-2005,Dolocan-2006,Moler-2005,Cliff-2010,Bending-2011}. For fields aligned along the $c$-axis scanning $\mu$-SQUID measurements on crystals with a $T_c=1.35$~K \cite{Dolocan-2005} indicated  `vortex coalescence'  at low fields, meaning that the vortices formed  densely-packed regions amid areas of low or zero vortex density.  
 Recently  scanning Hall probe microscopy  has been used to demonstrate that vortex behaviour can be significantly influenced by small differences in sample purity  \cite{Bending-2011}. In the best samples a change was observed in  local vortex correlations from triangular to square symmetry as the field was increased.
 
 Theoretical calculations of the vortex lattice have been performed using a Ginzburg-Landau (GL) approach  based  on a two-dimensional odd-parity superconducting order parameter with chiral $p$-wave symmetry \cite{Heeb-99}.  Four-fold symmetric vortices were predicted that were shown, using an extended London theory, to form triangular coordinated structures close to $H_{c1}$ that continuously deform into a square lattice as the field is increased. 
  More recently attempts have been made to explain the vortex coalescence observed experimentally \cite{Dolocan-2005,Dolocan-2006,Cliff-2010}.  It is proposed that this could be realistically explained in terms of multiple orbital degrees of freedom and the multiband superconductivity \cite{Egor1_2011,Egor2_2011,Garaud}.  Multiple effective coherence lengths lead to a semi-Meissner state   whereby a combination of  long-range attractive and short-range repulsive  inter-vortex interactions give rise to  clusters of vortices nucleating within a  Meissner-like state \cite{Egor_2005}.  
 This behaviour was termed Ôtype-1.5Õ superconductivity when reported in MgB$_2$ \cite{Moshchalko-10,Chibotaru-11,Gutierrez-12}. Non-pairwise interactions may also lead to tendancies to form chain-like or irregular clusters \cite{Egor2_2011} not unlike those observed in some surface probe experiments \cite{Dolocan-2006,Moler-2005}.

In this paper we use $\mu$SR to probe the {\em bulk} superconducting state of Sr$_2$RuO$_4$, where significant differences to {\em surface}  probe measurements are found \cite{Bending-2011}.  At low fields and temperatures we observe the striking phase separation of the sample into regions containing vortices, embedded within significantly larger regions from which the flux is completely excluded.  This  coexistence of mixed state and Meissner-like  regions has been predicted for Sr$_2$RuO$_4$,  where such  a semi-Meissner state emerges within a type-1.5 scenario \cite{Egor1_2011,Egor2_2011,Garaud}. In addition we observe a  triangular to square transitions with both increasing field and temperature, as predicted by models of $p$-wave or chiral $p$-wave type-1.5 superconductivity \cite{Heeb-99,Garaud}.

\section{Experimental Method}
The  Sr$_2$RuO$_4$ single crystals  were grown using the floating zone technique with Ru self-flux in an image furnace.  The samples were annealed in air for 3 days at 1500$^{\circ}$C in order to reduce lattice defects. The superconducting transition temperatures were determined using $ac$-susceptibility, with  $T_c=1.5$~K. The $\mu$SR sample consisted of a mosaic of crystals mounted onto a silver backing plate in a dilution refrigerator with their $c$-axes  aligned perpendicular to the  plate and parallel to the momentum of the incoming muons.  All measurements were made after cooling the samples from above $T_c$ in an applied field.
 
In a transverse $\mu$SR experiment spin polarised muons are rapidly  brought to rest inside the sample where they precess about the local internal flux density $B$ at an angular frequency $\omega = \gamma_\mu B$ determined by the gyromagnetic ratio of the muon $\gamma_\mu = 851$ MRads s$^{-1}$T$^{-1}$. The muons decay with an average lifetime of 2.197~$\mu$s, emitting positrons preferentially along the muon-spin direction.  The difference between two positron detectors placed at opposite edges  of the sample can be used to obtain the time evolution  of the decay asymmetry $\mathcal{A}(t)$. In the current experiments an external magnetic field is directed parallel to the $c$-axes, so that a component of the spin-polarisation precesses in a plane parallel to the $ab$-planes of the crystals. $\mathcal{A}(t)$  thus samples the density variation within the $ab$-planes of the component of flux  parallel to the $c$-axis.   A  maximum entropy technique \cite{Riseman}  was used to Fourier transform  $\mathcal{A}(t)$ in order to produce the probability of internal flux density $p(B)$ as discussed in ref.~\cite{Sonier}.  The form of $p(B)$ is strongly related to the spatial correlations of  the vortex lattice \cite{Heron-2013}. To acquire data of sufficient quality for this analysis at least 2 hours of data collection are required  for each sample condition (temperature, field), so that the measurements average over this period.  For a given condition there was no evidence for changes of the magnetic state of the sample as a function of time, as would seem reasonable for field-cooled measurements.

\section{Results}
Theoretical treatments of the exotic superconducting state in Sr$_2$RuO$_4$ make definite predictions about the evolution of vortex lattice symmetry with magnetic field and temperature. We first discuss what our measurements are able to reveal concerning the vortex lattice symmetry and perfection, before proceeding to discuss  direct signatures of  type-1.5 superconductivity via the observation of vortices in the semi-Meissner state.

\begin{figure}[h]
\begin{center}
\includegraphics[width=14cm]{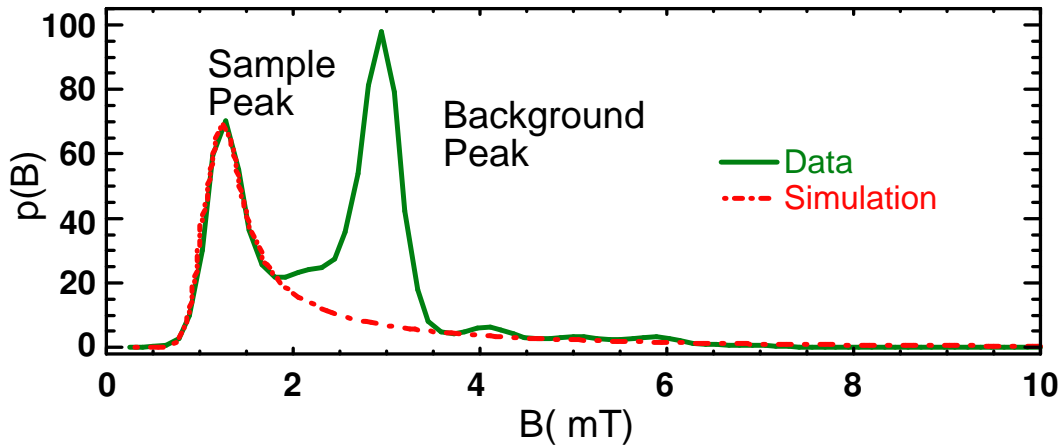}
  \caption{\small Example of  a $\mu$SR lineshape $p(B)$ (solid green  line) taken at 2.8 mT, 500 mK showing a simulation of  a triangular structure (dashed red).  The second (unmodelled) peak is the background peak from the sample holder, which at this temperature is large relative to the superconducting signal (see text).  }
 \label{fig:lineshapes}
 \end{center} 
 \end{figure}
 
 An example of an experimental lineshape $p(B)$ is given in Fig.~\ref{fig:lineshapes} taken after field-cooling from above $T_c$ to 500 mK in an applied field of  2.8~mT.
The dashed line represents a numerical simulation of the contribution to the data (solid line) arising from the vortex lattice. The additional peak in the data represents the background signal from muons stopping not in the sample but in the backing plate and other parts of the cryostat within the area of the muon-beam. The absolute magnitude of the background was found to be independent of applied field and temperature.  The large size of the background relative to the signal reflects the reduction, at low field and low temperature, of the component of the signal arising from the vortex lattice, discussed later.

The general form of the $p(B)$ associated with a periodic lattice can be seen in the simulation (Fig.~\ref{fig:lineshapes}).
The data were modelled by a real-space numerical simulation, calculating the $p(B)$ for the $B(r)$ within a unit cell embedded within a large vortex lattice.
A real space simulation was chosen so that in addition to simple triangular and square lattices a variety of other ordered and disordered systems could easily be explored, such as disordered chain structures. The model is able to continuously alter the angle $\theta$ between the lattice vectors $\vec{a}_1$, $\vec{a}_2$,  the length of which can also be varied to create, for example, tetragonal unit cells.  The vortices were modelled as London-like  by a second order modified Bessel function with cores that were included {\em via} a smoothing of the  singularity over a characteristic length scale $\xi_o$.  For these low-field data the form of  the core has little effect on the simulations, which are sufficient to capture the essential features associated with symmetry and perfection of the lattice.  All lineshapes were also convoluted with a Gaussian  of width $\sim 0.2$~mT   to represent the effects of instrumental broadening.  In Fig.~\ref{fig:lineshapes} (2.8 mT, 500 mK) the data have been modelled using an isotropic triangular lattice, with the best description of the data being given using a penetration depth  $\lambda_{ab}\sim 200$~nm.
 Significant anisotropy, if present, would manifest as a double-peaked structure in $p(B)$ due to the occurrence of two  saddle points in the unit cell. For these data  no such splitting  is resolved. If present such anisotropic distortions must thus be very small. For this reason all simulations in this paper are therefore of isotropic lattices,  although other possibilities were explored.   
 
  \begin{figure}[h]
\begin{center}
 \includegraphics[width=14cm]{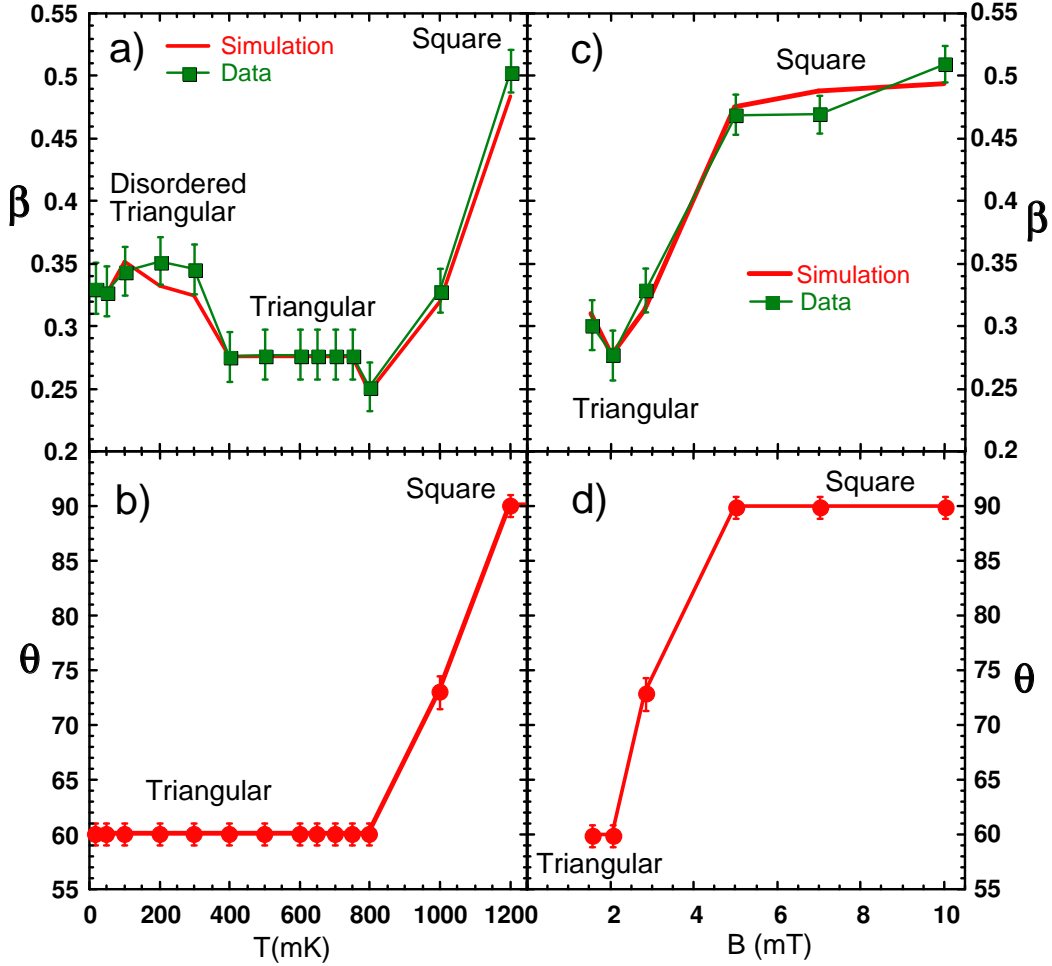}
  \caption{\small a-b: Temperature-dependent parameters for data taken  at 2.8 mT.  a) The parameter $\beta$ (red line: simulation, green squares: data) indicating both the change of symmetry and the onset of stronger positional fluctuations below 400 mK. b) The angle $\theta$ between the vortex lattice vectors $\vec{a}_1$, $\vec{a}_2$ used in our simulations of the data.  c-d: Parameters for field-cooled data taken at 1~K as a function of applied field.  c) The parameter $\beta$ (red line: simulation, green squares: data) indicating the crossover to a square lattice above $\sim$~3mT.  d) The angle $\theta$.}
 \label{fig:theta-and-beta}
 \end{center} 
 \end{figure}

 \begin{figure}[h]
\begin{center}
\includegraphics[width=14cm]{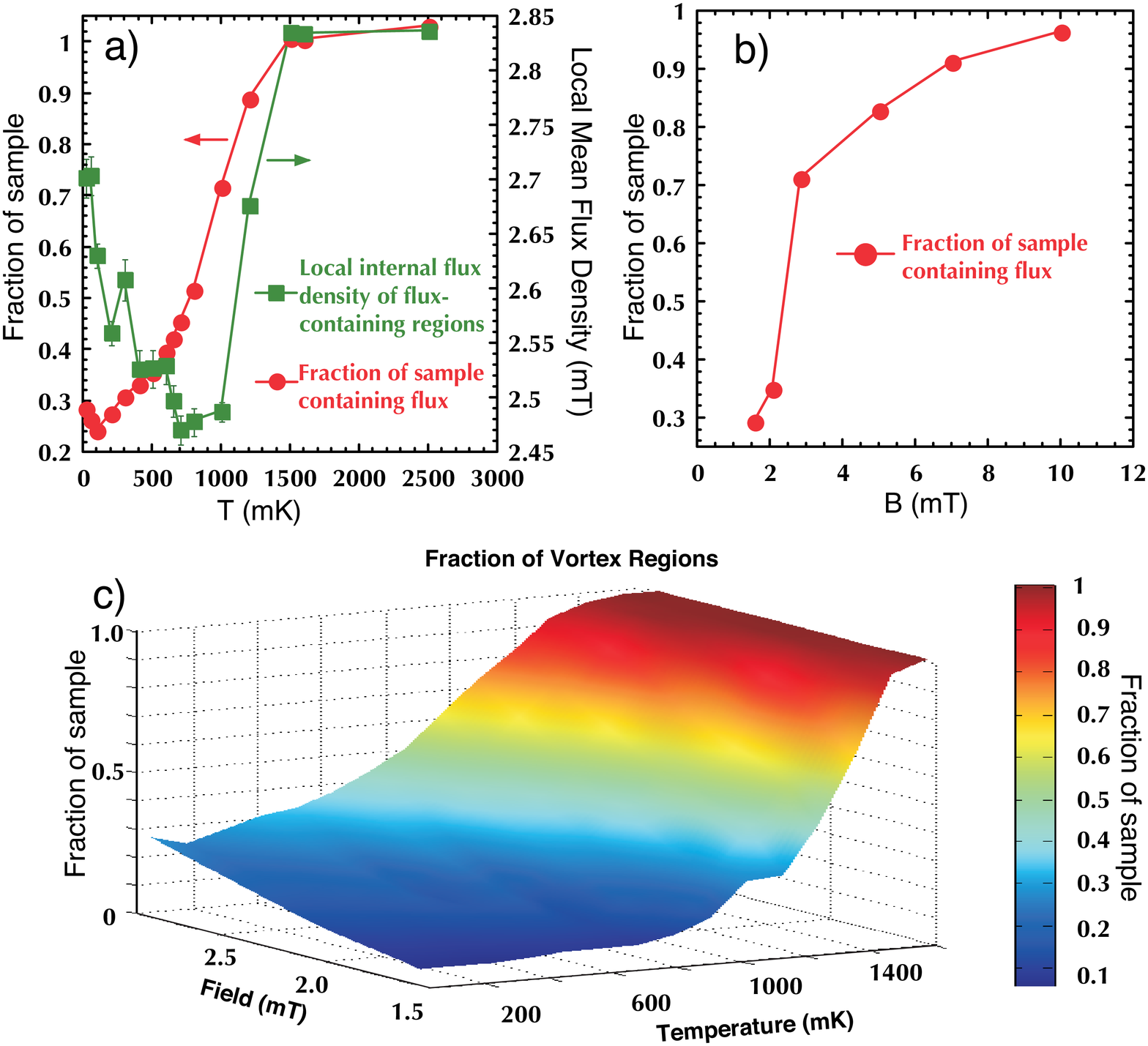}
  \caption{\small The fraction of  the sample containing flux, derived from the measured asymmetry.  a) As a function of temperature at 2.8~mT, b) As a function of field at 1K, indicating the reduction of volume of  flux-containing regions. The rise in the mean local flux density within these regions with decreasing temperature is also shown in (a). c) Low-field evolution of the fraction of vortex-containing regions for $T<T_c$.}
 \label{fig:asymm}
 \end{center} 
 \end{figure} 
 
A useful way to parameterise the superconducting contributions to the lineshapes is to calculate the quantity \cite{Lee-97}
$\beta = \bigg(\frac{B_{pk}\,-\,B_{min}}{\sigma}\bigg)$,
where $B_{pk}$, $B_{min}$ and $\sigma$ are the mode field, minimum field and root second moment of the superconducting contribution to the $p(B)$.  For ideal lattices $\beta$ provides a measure of the symmetry, changing from around $\beta \sim 0.3$ for a triangular lattice to  $\beta \sim 0.5$ for a square lattice.  
In Fig.~\ref{fig:theta-and-beta}a  $\beta$ is plotted as a function of temperature, derived both from the data at 2.8~mT and also from the simulations of the data using our model.    
In Fig.~\ref{fig:theta-and-beta}b we plot the corresponding  angle $\theta$ determined from simulations of the data, which exhibits a cross-over  from $\theta =60^{\circ}$ (triangular) to $\theta =90^{\circ}$ (square) beginning above 800~mK.
This cross-over is also reflected in the value of $\beta$ in  both the data and the simulations, which increases rapidly above  $800$~mK.   For non-ideal lattices $\beta$ also reflects the influence of  disorder, modelled as random positional fluctuations $\langle u^2\rangle^{\frac{1}{2}}$ about the ideal positions.  At 2.8~mT  these are found to be very small for temperatures down to 400~mK, but  rise to values of $\langle u^2\rangle^{\frac{1}{2}}/a\sim5\%$ below this temperature, where $a\sim\sqrt{\Phi_o/B}$ is the lattice parameter.   This low temperature region  is thus characterised as a lattice with an underlying triangular coordination but with a significant amount of random positional disorder.  

It is also possible to use $\beta$, as well as $\theta$ derived from simulations to the data, to quantify the evolution of the lattice symmetry with increasing field. At a temperature of 1~K ((Fig.~\ref{fig:theta-and-beta}c \& d), the low-field lineshapes are characterised by values of $\beta$ and $\theta$ indicating a triangular lattice, crossing over completely to a square lattice by  5 mT.   The observation of a square lattice at 5~mT and above is in agreement with previous measurements using $\mu$SR \cite{Aegeter-98} and small-angle neutron scattering \cite{Riseman-98}.  The transition to a triangular lattice at low field is also in accord with the scanning Hall probe measurements \cite{Bending-2011}.

\begin{figure}[h]
\begin{center}
 \includegraphics[width=14cm]{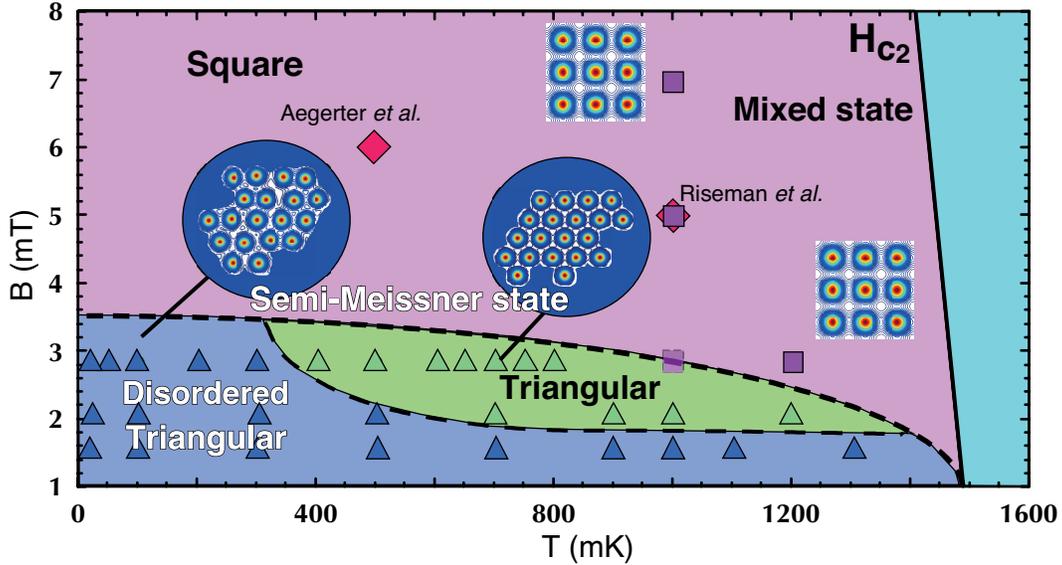}
  \caption{\small a) A phase diagram based on the $\mu$SR measurements  broadly indicating the  areas of vortex lattice symmetry and perfection. Symbols indicate measurement points: square lattice (squares); well-ordered triangular lattice (light, green triangles); disordered  triangular lattice (dark, blue triangles); square lattice refs.\cite{Riseman-98, Aegeter-98} (red diamonds). 
}
 \label{fig:phasediagram}
 \end{center} 
 \end{figure}

We now examine the most remarkable and important feature of our data, the striking reduction in the volume of sample occupied by vortices as the field and temperature are reduced. 
The measured initial asymmetry $\mathcal{A}_0$ reflects the volume of the sample in which muons are measured to precess. While the contribution to $\mathcal{A}_0$ from the background signal always remains constant, the  total measured asymmetry {\em falls dramatically} and {\em continuously} as a function of decreasing field and temperature.  This is illustrated  in Fig.~\ref{fig:asymm}, showing the fraction of the total volume of the sample where muons are able to precess, which is strongly reduced  as a function of both decreasing  applied field and decreasing  temperature. The formation of a uniform vortex lattice should not lead to any change of $\mathcal{A}_0$ for  $T<T_c$, whereas  the formation of a Meissner state should lead to a state with $\mathcal{A}_0=0$ in the {\em precessing}  component.  In principle in zero flux density, atomic and nuclear magnetic moments  lead to a very slow {\em  relaxation}   of the muon polarisation; in these very pure samples our measurements within the normal state indicate a slow relaxation rate of 0.02$\mu$s$^{-1}$ or less in this temperature range. The contribution of this to the maximum entropy Fourier transforms  such as those in Fig.~\ref{fig:lineshapes}, which are located around the applied field, is effectively zero.  Experimentally we  thus measure a reduction of asymmetry in the {\em precessing} signal,  indicating an increasing volume of sample from which the applied magnetic field is {\em completely excluded} (no precession), at the expense of vortex containing regions that diminish in volume (~25\% at 2.8~mT, 100~mK).
In very low $\kappa$ materials demagnetisation effects can cause an intermediate state to form in which regions of Meissner state coexist with regions of {\em homogeneous}  flux penetration, the latter characterised by internal fields much {\em greater} than the applied field. $\mu$SR can directly probe these states, even revealing (e.g. in LaNiSn) the coexistence of small pockets of mixed state with fields above the value of the applied field \cite{Drew}. By contrast, what we observe in Sr$_2$RuO$_4$ is qualitatively very different. We observe {\em no intermediate state}, just a diminishing vortex fraction and a growing Meissner fraction, so that both coexist.
 
 \section{Discussion}
The semi-Meissner state we observe  is precisely that predicted  by theoretical models  of type-1.5  superconductivity suggested by the multiband superconductivity of Sr$_2$RuO$_4$ \cite{Garaud, Egor1_2011,Egor2_2011}. Long-range attractive forces lead to the formation of a semi-Meissner phase comprising vortex clusters embedded within a Meissner-like matrix,  the formation of which $\mu$SR experiments are uniquely able to follow in the bulk.  Further key evidence for this interpretation can be seen in Fig.~\ref{fig:asymm}a.   Below $T_c$,  as expected the mean internal flux density within the vortex-containing regions initially begins to fall. Below  $\sim 500$~mK, however, the flux density within these regions, and hence the density of vortices, begins to {\em increase} again, giving strong evidence for the onset of a {\em  long-range vortex attraction}.  At this field the vortex separation within these clusters is large, $d\sim 4 \lambda_{ab}$, which, combined with the fact that the vortex separation reduces only when $T\ll T_c$, is highly consistent with a type-1.5 scenario with a passive band having relatively small superfluid density  \cite{Egor1_2011}.   At lower applied fields this upturn in the density at low temperature is no longer observed, suggesting that  at temperatures where long-range attraction sets in  the vortex system is too dilute to form large clusters.  At 1.6~mT ($d\sim 6 \lambda_{ab}$) the sample is almost entirely in a Meissner state (Fig.~\ref{fig:asymm}c), with only a tiny signature from the mixed phase remaining.  This argues strongly against local inhomogeneity as the source of cluster formation. The vortex coalescence observed in surface probe experiments  \cite{Dolocan-2005,Dolocan-2006,Cliff-2010} may be related to our striking observations, although in those works alternative explanations,  including pinning inhomogeneity or small misalignment of the field and the $c$-axis,  could not  be ruled out.  

Despite the excellent qualitative agreement  with multiband type-1.5 models, one cannot entirely rule out that what we observe may also be influenced by the existence of the putative chiral order parameter in Sr$_2$RuO$_4$.   In zero field, domains of opposite chirality can form below $T_c$. If a sample has multiple chiral domains then  the application of
a small external field can cause the nucleation of small vortex regions within the domains of the preferred chirality, which grow in area with increasing field, as has been modelled in ref.~\cite{Ichioka_2012}. This interpretation seems less likely, however, given that:
(i) all our measurements are slowly field cooled  and the symmetry is thus already broken as the sample cools through $T_c$; (ii) the magnetic field in the clusters increases substantially with decreasing temperature; and (iii) there is a significant decrease in the tendency to cluster in low fields.
 
In recent scanning Hall probe experiments carried out on samples from the same source as those used in the $\mu$SR experiments,  no significant evidence was reported for large flux-free regions, though the observation of both triangular and square vortex correlations was in reasonable agreement \cite{Bending-2011}. This might be partially explained if the vortices splay and exit the sample over a large fraction of the surface, while threading  a smaller fraction of the bulk volume \cite{Garaud}. This may be a contributory factor to the anomalous broadening of the flux line profile observed in ref.~\cite{Bending-2011}. More generally we emphasise that $\mu$SR is a {\em bulk} technique and may thus probe different aspects of the superconducting state to surface techniques, which may need to be reconciled theoretically.

Focussing on the vortex-containing regions, the data  and lineshape analysis, summarised in terms of $\beta$ and $\theta$, allows us to construct the phase diagram shown in Fig.~\ref{fig:phasediagram}.  In the low-field low-temperature corner, where there is evidence of disorder, we explored a number scenarios but find the data to be most consistent with an isotropic disordered  triangular lattice. This disorder may result  from a number of sources including competing  vortex lattice symmetries, finite cluster sizes, non-pairwise interactions \cite{Egor2_2011} and the  rearrangement of the vortices over  time scales longer than the measurement. This disorder is most pronounced at 2.8~mT,  the region where the increasing vortex density at low temperature is also most pronounced, but is also present to a lesser degree at lower fields. 

\section{Conclusions}
In conclusion we have explored the superconducting state within the bulk of high-quality Sr$_2$RuO$_4$. At low field and temperature we observe a phase-separated  semi-Meissner state, consistent with the nucleation of vortex clusters in type-1.5 scenarios arising from the multiband nature of the superconductivity. Perhaps the strongest evidence in favor of a type-1.5 regime is the unique observation of increasing flux density in the vortex clusters with decreasing temperature.  This is in agreement with calculations in the regime with multiple coherence lengths arising from active and passive bands, such that $\xi_1<\sqrt{2}\lambda < \xi_2$ \cite{ Egor1_2011}. In that case the attractive intervortex interaction can appear far below $T_c$, with the energetically preferred intervortex distance substantially diminishing with decreasing temperature. We cannot, however, completely exclude the possible influence of a chiral order parameter in nucleating small vortex regions, an effect that can naturally coexist with the physics of the type-1.5 state. We also present clear  evidence for a triangular to square cross-over as a function of increasing field and temperature. The quantitative measurement of the evolution with  field and temperature,  of both vortex fraction and changes of vortex lattice symmetry, presents a stimulus and a challenge to realistic models of superconductivity in this and related materials.

\acknowledgments
We acknowledge the financial support of the EPSRC (grant EP/J01060X).  All $\mu$SR experiments were carried out courtesy of the Paul Scherrer Institute. E. Babaev was supported by the US NSF CAREER Award No. DMR-0955902  and
by the Knut and Alice Wallenberg Foundation through the Royal Swedish Academy of Sciences, Swedish Research Council.


\bibliographystyle{unsrt}

\end{document}